\documentclass[useAMS,usenatbib]{mn2e}
\usepackage{psfig, epsf, epsfig}

\title[On the weakness of disc  
models in bright ULXs]{On the weakness of disc  
models in bright ULXs}

\author[A. C. Gon\c{c}alves \& R. Soria]
{A. C. Gon\c{c}alves$^{1,2}$\thanks{E-mail:
anabela.goncalves@obspm.fr} and R. Soria$^{3,4}$\thanks{E-mail:
rsoria@cfa.harvard.edu}\\
$^{1}$\/LUTH, Observatoire de Paris-Meudon, 5 Place Jules Janssen, 
       92195 Meudon, France\\
$^{2}$\/CAAUL, Observat\'orio Astron\'omico de Lisboa,
Tapada da Ajuda, 1349-018 Lisboa, Portugal\\
$^{3}$\/Harvard-Smithsonian Center for Astrophysics, 
	60 Garden st, Cambridge, MA 02138, USA\\
$^{4}$\/Mullard Space Science Laboratory (UCL), Holmbury St Mary, 
	Dorking, Surrey, RH5 6NT, UK}

\begin{document}

\date{Accepted \today }

\pagerange{\pageref{firstpage}--\pageref{lastpage}} \pubyear{2006}

\maketitle

\label{firstpage}

\begin{abstract}
It is sometimes suggested that phenomenological power-law 
plus cool disc-blackbody models represent the simplest, 
most robust interpretation of the X-ray spectra of bright 
ultraluminous X-ray sources (ULXs); this has been taken as evidence 
for the presence of intermediate-mass black holes (BHs) 
($M \sim 10^3 M_{\odot}$) in those sources. 
Here, we assess this claim by comparing the cool disc-blackbody 
model with a range of other models. For example, we show that
the same ULX spectra can be fitted equally well by subtracting
a disc-blackbody component from a dominant power-law component,
thus turning a soft excess into a soft deficit. Then, we propose
a more complex physical model, based on a power-law 
component slightly modified at various energies 
by smeared emission and absorption lines from highly-ionized, 
fast-moving gas. We use the {\it XMM-Newton}/EPIC spectra 
of two ULXs in Holmberg II and NGC\,4559 as examples.
Our main conclusion is that the presence  
of a soft excess or a soft deficit 
depends on the energy range over which we choose 
to fit the ``true'' power-law continuum; those small 
deviations from the power-law spectrum are well modelled  
by disc-blackbody components (either in emission or absorption) 
simply because they are a versatile fitting tool for most kinds 
of smooth, broad bumps. Hence, we argue that 
those components should not be taken as evidence  
for accretion disc emission, nor used to infer BH masses. 
Finally, we speculate that bright ULXs could 
be in a spectral state similar to (or an extension of) 
the steep-power-law state of Galactic BH candidates, 
in which the disc is now completely comptonized and 
not directly detectable, and the power-law 
emission may be modified by the surrounding, 
fast-moving, ionized gas. 
\end{abstract}


\section{Introduction: how to determine BH masses}

Ultra-luminous X-ray sources (ULXs) are point-like, 
accreting X-ray sources with apparent isotropic  
luminosities spanning a range from  $\sim 10^{39}$  
to $\approx 3 \times 10^{40}$ erg s$^{-1}$, 
that is, one or two orders of magnitude
greater than the Eddington luminosity ($L_{\rm Edd}$) 
of a stellar-mass 
black hole (BH). The main unsolved issue is whether 
the accreting sources are more massive than typical 
Galactic BH candidates (BHCs), perhaps in the intermediate-mass 
range ($M \sim 10^3 M_{\odot}$; Miller, Fabian \& Miller 2004), 
or stellar-mass BHs accreting at super-Eddington rates 
(Begelman 2002); alternatively, their brightness could 
be due to beaming along the line-of-sight of the 
observer (King et al.~2001; K\"{o}rding, 
Falcke \& Markoff~2002; Fabrika \& Mescheryakov 2001). 

The standard, most reliable way to determine the mass 
of an accreting BH in X-ray binaries is based 
on phase-resolved spectroscopic and photometric studies 
of their optical counterparts. By measuring the orbital 
period and the radial velocity shifts of selected optical 
lines from the donor star and the accretion disc, one 
can constrain the mass functions of both components. 
Further constraints to the inclination angle and the size 
of the system come from the amplitude of the ellipsoidal 
variations in the donor star, and the presence/absence 
of dips and eclipses. Those techniques have been 
successfully applied to a growing number of BHCs 
(for a review, see McClintock \& Remillard 2006). 
Attempts to apply similar techniques to ULXs have 
been fruitless or inconclusive, so far, mostly because  
of their optical faintness. At typical distances of a few Mpc 
(distance moduli $\sim 28$--$30$), most candidate 
optical counterparts are fainter than $V \sim 24$ mag. 
In many cases, crowding is also a problem: 
the X-ray error circle may be consistent with 
an unresolved group of stars. Pioneering efforts 
(e.g., Gris\'{e}, Pakull \& Motch 2006; 
Pakull, Gris\'{e} \& Motch 2006) may yield results 
for a few sources in the near future. Meanwhile, though, 
one has to rely on indirect methods to estimate the BH mass.

One such method is based on X-ray spectral fitting 
over the ``standard'' $0.3$--$10$ keV band. 
In Galactic BHCs, the X-ray spectrum 
consists of essentially two components, power-law 
and thermal, with varying 
normalizations and relative contributions in various 
spectral states. The power-law component is scale-free 
and without a direct dependence on BH mass. However, 
its slope and normalization are related to the spectral 
state and normalized luminosity; for instance,  
the slope is flatter (photon index $\Gamma \sim 1.5$--$2$) 
in the low/hard state ($L_{\rm X}/L_{\rm Edd} \la 10^{-2}$) 
and steeper ($\Gamma \sim 2.5$--$3$) at 
$0.1 \la L_{\rm X}/L_{\rm Edd} \la 1$ (McClintock \& Remillard 2006). 
More significantly, the thermal component, interpreted 
as the spectrum of an optically-thick Shakura-Sunyaev disc 
(Shakura \& Sunyaev 1973), 
contains, in principle, a direct dependence on 
disc size and BH mass.

In the standard thin-disc approximation, for 
a non-rotating BH, 
\begin{equation}
\sigma T_{\rm eff}^4(R) 
\approx \frac{3 G M \dot{M}}{8 \pi R^3} 
\end{equation} 
(Shakura \& Sunyaev 1973; Pringle 1981; 
Frank, King \& Raine 2002)\footnote{Strictly speaking, 
in the standard disc-blackbody approximation the disc  
temperature quickly drops to zero at the inner edge, 
due to the zero-torque condition at that boundary, 
after reaching a maximum at $R \approx (49/36) R_{\rm in}$. 
However, equation (1) corresponds to the simplified version 
of the disc-blackbody spectrum more commonly used 
in data fitting ({\tt diskbb} model: Makishima et al.~1986), 
in which the temperature is assumed to increase 
all the way to the innermost stable circular orbit, $R_{\rm in}$. This is why 
the fitted peak temperature of a disc-blackbody model 
is commonly referred to as the inner-disc temperature, $T_{\rm in}$.}. 
If the disc extends 
to the innermost stable circular orbit, 
$R_{\rm in} = 6GM/c^2$ in the Schwarzschild geometry, 
the inner-disc temperature scales as 
\begin{equation}
T_{\rm in} \equiv T(R_{\rm in}) \propto M^{-1/2} \dot{M}^{1/4}. 
\end{equation} 
The total luminosity of a standard disc, 
mostly emitted in the X-ray band for stellar-mass 
systems, is (Makishima et al 1986) 
\begin{equation}
L_{\rm X} \approx 4\pi \sigma T_{\rm in}^4 R_{\rm in}^2 
\approx 9.76 \times 10^{36} 
\left(\frac{T_{\rm in}}{{\rm keV}}\right)^4 
\left(\frac{M}{M_{\odot}}\right)^2  ~~{\rm erg~s}^{-1}~ 
\end{equation}
which implies that $L_{\rm X} \propto T_{\rm in}^4$, 
varying the accretion rate along lines of constant BH mass. 
At the Eddington limit, assuming a standard radiative 
efficiency $\approx 0.1$, 
\begin{equation}
\dot{M}_{\rm Edd} \approx \frac{1.3 \times 10^{39}}{c^2} \, 
\left(\frac{M}{M_{\odot}}\right) ~~{\rm g~s}^{-1},
\end{equation}
therefore $T_{\rm in} \propto M^{-1/4}$ from Eq.~(2), and 
\begin{equation}
L_{\rm Edd} \approx 1.3 \times 10^{38} 
\left(\frac{M}{M_{\odot}}\right) 
\approx 4.5 \times 10^{39} 
\left(\frac{T_{\rm in}}{{\rm keV}}\right)^{-4} ~~{\rm erg~s}^{-1}.~
\end{equation}

There are various caveats attached 
to those simple relations. For example, the innermost stable circular 
orbit (assumed to mark the inner-disc boundary 
at high accretion rates) can vary between 
$6GM/c^2$ and $GM/c^2$ for a non-rotating or maximally-rotating 
BH, respectively. The fitted temperature (``color temperature'') 
may be higher than the effective temperature 
by a factor as high as $2.6$ (Shrader \& Titarchuk 2003).
On the other hand, this is approximately compensated 
(Fabian, Ross \& Miller 2004) by the temperature 
drop near the inner edge, due to the zero-torque condition.
The normalization of the observed spectrum, 
and therefore also the inferred BH mass, depend 
on the viewing angle, often poorly constrained.

Nevertheless, equations (3) and (5) provide an overall  
good scaling and order-of magnitude estimate of the BH mass 
in Galactic BHCs, and the {\tt diskbb} model in {\small XSPEC} 
(Arnaud 1996) has proved simple and successful. 
More complex implementations of the disc-blackbody 
model (e.g., {\tt diskpn}; Gierlinski et al.~1999), 
taking into account some of the corrective factors mentioned above,
have also been used, at the cost of additional free parameters. 
It seems reasonable to apply the same simple tools 
to estimate the mass of the accreting BHs 
in ULXs, if they are scaled-up versions of Galactic 
BHCs.

\section{Competing models}


\subsection{Cool\,disc\,phenomenological\,vs.\,physical\,models}

For about a dozen of the brightest ULXs, it was noted 
(e.g. Miller et al.~2003; Miller, Fabian \& Miller 2004; 
Fabian, Ross \& Miller 2004; Roberts et al.~2004; 
Terashima \& Wilson 2004; Feng \& Kaaret 2005) 
that the $0.3$--$10$ keV spectrum is dominated by  
a featureless broad-band component, interpreted as a power-law 
({\tt {po}} in {\small {XSPEC}}) 
plus a ``soft-excess'' significantly detected below 1 keV. 
The normalization of the soft excess is somewhat 
degenerate with the column density, metal abundance 
and ionization state of the absorbing medium. However, 
for various sources there is little doubt that an additional 
thermal component ({\tt{bb}} or {\tt {diskbb}} 
in {\small {XSPEC}}) with $kT \sim 0.1$--$0.2$ keV
leads to better fits. 
 
By analogy with Galactic BHCs, 
one can interpret such phenomenological fits  
as true physical models and use the fitted 
temperature as the color temperature near the inner boundary of 
an accretion disc; by applying equation (3), we thus obtain  
characteristic mass values $\sim 10^{3} M_{\odot}$ 
(Figure 1). This approach has the advantage
of being simple, with a minimum number of free parameters, 
well tested for Galactic BHCs, and easy to apply as ``common currency'' 
(Miller, Fabian \& Miller 2006) in the comparison 
and classification of different sources.
It does not assume or require a specific physical model 
for the power-law component. 

An alternative approach, also successfully applied to Galactic 
BHCs, 
is to fit the X-ray spectra with a more complex, self-consistent 
model (e.g., {\small {XSPEC}} models such as 
{\tt eqpair}: Coppi 1999; {\tt bmc}: Shrader \& Titarchuk 1999; 
{\tt comptt}: Titarchuk 1994), in which a power-law-like component 
arises as comptonized 
emission from seed thermal photons, upscattered in a corona.
In this class of physical models, the thermal component 
(which we can still interpret as disc emission) 
is slightly modified from the pure {\tt diskbb} spectrum, 
and the power-law has a low-energy truncation at the energy 
of the seed photons, and a high-energy break 
depending on the temperature of the corona. 
The temperature of the seed thermal component, 
and the optical depth and temperature of the comptonizing corona 
are the main fitting parameters.
When such models are applied to bright ULXs, 
it is found (Goad et al. 2006; Stobbart, Roberts \& Wilms 2006) 
that the emission from the inner disc may be almost 
completely comptonized in a warm, optically-thick, 
but perhaps patchy, corona.

This alternative approach was critically discussed by  
Miller, Fabian \& Miller (2006), who pointed out 
that complex comptonization models contain a larger 
number of additional parameters that can be adjusted
to fit the data.   Miller et al.~(2006) 
simulated a {\tt diskbb} $+$ {\tt po} spectrum 
and pointed out that it could also be fitted 
with complex comptonization models. 
In some cases, the same simulated spectrum could be fitted 
with different sets of physical parameters: either 
an optically-thin, hot corona ($\tau \approx 0.8$, 
$kT_e \approx 50$ keV), or a thicker, warmer one ($\tau \approx 4$, 
$kT_e \approx 7$ keV).
Thus, Miller et al.~(2006) concluded that, because of  
the moderately low signal-to-noise ratio and few distinctive 
features in ULX spectra, comptonization models cannot give us 
a more solid understanding of the physical situation 
than simpler phenomenological models.

However, we do believe that complex physical 
models can be useful to draw our attention to various, 
apparently minor issues, which are not considered in simpler models, 
but can significantly affect our physical interpretation.
For example, the temperature of the warm corona 
fitted to the spectrum of Holmberg II X-1 (Goad et al.~2006)
is only $\approx 3$ keV, which suggests 
a steepening in the spectral slope 
at energies $\ga 5$\,keV. A similar spectral 
steepening has also been found to be statistically 
significant for most of the ULXs with higher 
signal-to-noise ratios (Stobbart et al.~2006).
Given the very small number of spectral features, 
such break can be an important 
clue to our understanding of the emission process.
A phenomenological {\tt diskbb} plus {\tt po} model 
simply cannot reproduce that break. 

We should also be careful when we attribute 
physical meaning to phenomenological fit parameters 
such as the colour temperature in the {\tt diskbb} model.
For example, it is possible that the inner disc 
may be cooler and fainter than a standard disc-blackbody, 
for a given mass and accretion rate.
This must happen not because the disc is truncated 
or replaced by 
an advection-dominated flow---in those cases, 
the total efficiency would also be low and the ULX 
would not be as bright, unless it was also very massive---but 
because most of the gravitational energy is transferred 
(e.g., via magnetic coupling) from the disc 
to the corona, before being radiated. In this scenario, 
the corona would draw energy from the disc 
and effectively cool it (e.g., see 
Kuncic \& Bicknell 2004 for a magnetized disc/corona 
model based on this idea).
There is no observational evidence yet 
to favour this scenario over the simpler 
{\tt diskbb} model. However, it is at least 
a reminder that an observed cooler disc 
may not necessarily be due to a higher mass.

\begin{figure}
\epsfig{figure=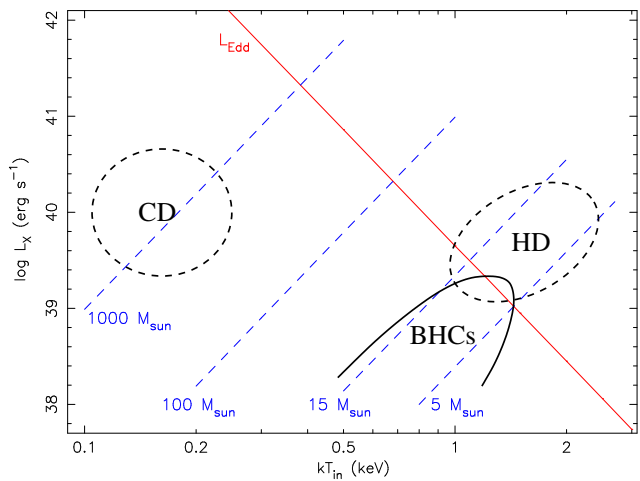, width=8.4cm, angle=0}
\caption{Schematic plot showing the location of 
Galactic BHCs and ULXs in a disc temperature versus 
X-ray luminosity plot; adapted from fig. 3 in Stobbart 
et al.~(2006). The Eddington limit and lines of constant 
mass are obtained from equations (3) and (5) (see Section 1 
for details). The CD model implies that ULXs are intermediate-mass 
BHs, emitting well below their Eddington limit. The HD model 
suggests that ULXs are stellar-mass objects (an extension 
of the Galactic BHC class), emitting above Eddington.}
\end{figure}

\subsection{Cool-disc\,vs\,hot-disc\,phenomenological\,models}
 
If it is misguided directly to compare phenomenological and complex 
physical models, it is instead fair to compare 
two phenomenological models, equally simple, with 
the same number of free parameters, 
but with entirely different physical interpretations.
Such a comparison can be done between the cool-disc model 
(Miller et al.~2004) and the hot-disc model (Stobbart et al. 2006).
As extensively discussed in Stobbart et al.~(2006), 
the latter scenario suggests that at least some bright ULXs 
may occupy a region of the parameter space adjacent 
to that occupied by Galactic BHCs (Figure 1). 
The smooth spectral component, dominating in the hard band, 
is modelled as disc emission, with temperatures 
$kT_{\rm in} \sim 1$--$2.5$ keV; the soft excess is well 
modelled by a simple blackbody component, and may be 
physically understood as downscattered emission, 
for example in an optically thick outflow 
(King \& Pounds 2003).
The cool-disc (CD) and the hot-disc (HD) models 
provide equally good fits to most ULXs. Here, 
we briefly discuss some of the strengths 
and weaknesses of the two scenarios.

\begin{itemize}
\item The CD model has been used as evidence 
of BH masses $\sim 10^3 M_{\odot}$, which
require more complicated (and so far untested) 
formation scenarios. Curiously, in the CD interpretation, 
ULX luminosities would remain always an order 
of magnitude below their Eddington limit (Figure 1), 
even for the brightest sources, suggesting perhaps 
some kind of upper limit to the mass supply. 
This behaviour is not observed 
in Galactic neutron stars and BHCs, which often 
reach or even slightly exceed their Eddington limit.
On the other hand, the HD model suggests that 
ULXs could be stellar-mass BHs emitting 
at up to an order of magnitude above their 
Eddington limit. From a physical point of view, 
the two scenarios 
(stellar-mass but well above Eddington, 
or $\sim 10^3 M_{\odot}$ but well below Eddington)
have different kinds of drawbacks, 
but more constraining observations in other 
energy bands will be necessary to rule either one out.

\item The standard relations between luminosity, inner-disc 
temperature and mass (summarized in Section 1) are well tested 
for systems such as Galactic BHCs in the high/soft state, 
when the disc contributes most of the emission. This would also 
be the case in the HD model, where the disc contributes 
$\sim 70$--$90\%$ of the X-ray emission (Stobbart et al.~2006). 
Conversely, in the CD scenario, the disc contributes  
only $\sim 5$--$20\%$ of the X-ray luminosity; hence, 
a direct scaling of the disc quantities from 
Galactic BHCs may not be appropriate.

\item In the HD model, the dominant spectral component 
is modelled with a simple {\tt diskbb} term, based on 
the standard, Shakura-Sunyaev disc. The fitted temperature 
and normalization imply super-Eddington luminosities.
However, at those luminosities, we do not expect the accretion flow  
to be consistent with the standard thin disc. In fact, 
we already see deviations from the standard-disc spectrum 
in Galactic BHCs as they approach their Eddington limit 
(Kubota \& Makishima 2004; Kubota \& Done 2004; 
see also Kubota, Done \& Makishima 2002 for an earlier 
{\it ASCA} study of a ULX). 
Taking into account those effects would require again 
complex physical models with various free parameters. 
If we want to stick to the simple 
{\tt diskbb} model for a phenomenological study, we need 
at least to be careful in attributing true physical meaning 
to its fit parameters.

\item The CD model cannot explain the observed break 
in the spectral slope at high energies (typically, 
going from a photon index $\Gamma \approx 2$ at 1.5--5\,keV, 
to a slope $\approx 3$ at 5--10\,keV), 
as noted earlier. The HD model does predict that break; 
in fact, if anything, it overpredicts it, because 
we expect the Wien tail of the thermal emission 
to drop exponentially just above the fitted energy 
range. Spectral observations in the $\sim$\,10--20\,keV 
region would provide a crucial test between 
the two phenomenological models.

\item The inner-disc temperature in the CD model 
is in the same range as the characteristic temperature 
of the soft excess in Seyfert 1s, despite the large 
mass difference between ULXs and AGN. Both kinds of 
spectra can formally be well fitted with a cool {\tt diskbb} 
component plus a power-law. However, attributing the same physical 
meaning (disc emission) to the soft excess in both systems 
is problematic, and would require ad hoc modifications, 
for example a strong color factor in AGN (thus, destroying 
the simplicity of the phenomenological model). 
More likely, the soft excess in AGN could be explained 
by a combination of blurred emission 
and absorption lines and reflection (Gierlinski \& Done 2004; 
Crummy et al.~2006; Chevallier et al.~2006). 
This is an example of how a simple, successful phenomenological 
model, such as the CD model, can lead to misleading or plainly 
wrong physical interpretations 
in at least one of those two classes of objects. 
On the other hand, characteristic disc temperatures 
in the HD model fall within, or close to, the range of stellar-mass BH 
temperatures.


\end{itemize}

\subsection{Slope of the power-law component}

Most of the discussion so far has focused
on the interpretation of the soft component. 
However, another unexplained finding of the CD model 
is that the slope of the dominant power-law is harder 
($1.5 < \Gamma < 2.5$; Stobbart et al.~2006; 
Roberts et al.~2005) than generally 
observed from bright Galactic BHCs in the high/soft 
or steep-power-law states 
($\Gamma > 2.5$; McClintock \& Remillard 2006).
This appears to be true both for some bright ULXs with a soft excess, 
and, even more so, for those whose spectra can be fitted with 
a simple power-law; for example, NGC 4559 X-2 
($L_{0.3-10} = 1.3 \times 10^{40}$ 
erg s$^{-1}$, with $\Gamma \approx 1.8$; Cropper et al.~2004)  
and M\,99 X-1 ($L_{0.3-10} = 1.6 \times 10^{40}$ 
erg s$^{-1}$, with $\Gamma \approx 1.7$; Soria \& Wong 2006, 
in preparation). More examples of ULXs with a power-law 
slope $\Gamma \la 2$ are discussed in Winter, 
Mushotzky \& Reynolds (2006), Terashima \& Wilson (2004), 
Roberts et al.~(2004).
One can take at least four alternative approaches to explain this finding.

1) It could be that such 
sources are intrinsically different from soft-excess ULXs 
because they are in the classical low/hard state 
(e.g., McClintock \& Remillard 2006),
characterized by a dominant, hard ($\Gamma \sim 1.5$--$2.1$) 
power-law, $0.3$--$10$ keV luminosities 
$\la 0.01 L_{\rm Edd}$, and a disc truncated 
at large distances from the BH.
Such a truncated disc would have a colour temperature 
below 100~eV (by analogy with stellar-mass BHs), 
thus rendering its observation with {\it {XMM-Newton}} or 
{\it {Chandra}} practicaly impossible.
If those ULXs are in the low/hard state, it would imply 
$L_{\rm Edd} \ga$ a few $10^{41}$ erg s$^{-1}$, 
corresponding to masses $\ga$ a few $10^3 M_{\odot}$.
It would support the intermediate-mass BH scenario, 
but again we should ask why we never see those sources 
reaching their Eddington limit. 

2) Alternatively, they could be in the high/soft state,  
with a standard disc extending down to the innermost 
stable orbit, and X-ray luminosities $\ga 0.1 L_{\rm Edd}$.
If so, we would expect that the phenomenological CD model 
should apply to both classes of sources: soft-excess and pure
power-law ULXs. Why then is the {\tt diskbb} component 
not detected in the latter class of ULXs?
Perhaps their BH masses are even higher 
($\ga 10^4 M_{\odot}$), thus pushing the disc component 
into the UV band, outside our detection band. This interpretation  
clearly carries with it many BH formation problems. 

3) Another, more likely possibility is that 
the {\tt diskbb} component is so tiny that it 
is undetectable at a given signal-to-noise ratio. 
As far as we know, a longer observation may turn 
a pure-power-law source into a soft-excess source. 
In other words, we have to accept 
that at any given signal-to-noise ratio, 
there is a number of bright ULXs  
in which the disc is not directly visible at all---perhaps 
because its emission is almost entirely comptonized, 
including that from the inner disc.
But this would undermine the possibility 
of inferring a mass from the fitted temperature
of the cool thermal component, when present: that 
component may simply be residual emission from 
the outer disc, while the hotter inner disc may not  
be directly visible. In this scenario, 
bright ULXs could be in an extreme form 
of steep-power-law state (McClintock \& Remillard 2006).
The discrepancy in the power-law slope between them and 
bright Galactic BHCs may be due to some missing 
details in our band-limited spectral fitting. For example, 
if ULXs and Galactic BHCs have different masses, 
we may be comparing their power-law slopes in two different 
energy ranges, in scaled units. Or, we may not 
be considering slight broad-band modifications to the power-law 
flux (e.g., reflection, or smeared emission/absorption lines)
that lead to a wrong estimate of the slope---for 
example, making it appear flatter in the $\sim 2$--$5$ keV range, 
in bright ULXs.

%

4) Finally, this unexplained slope of power-law continuum 
may be telling us that spectral state classifications 
in Galactic BHCs and ULXs are totally different. In that case, 
phenomenological models tested for Galactic BHCs may not 
be directly applicable, or may not have the same 
physical interpretation in ULXs.

\subsection{Soft excess or soft deficit?}

Both the phenomenological CD and HD models 
share the same bias: namely, that the dominant 
component of the spectrum is well determined  
by the observed emission at $\sim 2$--$5$ keV. For example, 
in the CD model, the spectrum is more or less 
a true-power-law in that energy range.
Deviations from the assumed true-power-law 
at energies $\la 2$ keV are thus cast in the form 
of a soft excess, while deviations at energies 
$\ga 5$ keV can be dismissed as small-count statistics, 
or with the introduction of an ad hoc cut-off, 
or by assuming a low-temperature corona 
if we are using a more complex comptonization model.
Similarly, the HD model assumes that the spectrum 
is a true disc-blackbody in that range, with its 
emission peak falling just below or around $5$ keV.
Again, this choice inevitably leads us 
to finding a soft excess below $2$ keV, modelled 
with an additional thermal component.
A possible reason for this bias is that the continuum 
in the $2$--$5$ keV spectral range is free from 
line-of-sight cold or warm absorption, hardly modified 
by any residual soft thermal-plasma emission, 
and has good spectral resolution and sensitivity 
in {\it Chandra} and {\it XMM-Newton}. 
Thus, if we use a power-law model, it appears natural 
to adjust its slope to fit that energy range, 
and then take care of any deviations.

Evidence for a change in the spectral slope 
in the $2$--$10$ keV band is given by Stobbart et al. (2006), 
who show that a broken power-law fit provides an improvement 
over a single power-law fit in 8 out of 13 ULXs 
in their sample (see their Tables 6 and 7). This supports the 
idea that most sources cannot be described by a single 
power-law continuum across the whole band. Thus, instead 
of estimating the continuum in the $2$--$5$ keV range, 
we could equally well assume that
the continuum in the region $\sim 5$--$10$ keV 
is the true expression of the power-law. 
If we do that, we find that most bright ULXs
have a distinctive ``soft deficit''.
We would then try to devise complex physical models 
to explain that deficit, or, more simply, 
we would use phenomenological 
models. By analogy with the CD model, where  
a {\tt diskbb} component is used to account 
for the smooth, broad-band soft excess, 
we could select a smooth, broad-band 
absorption component. In fact, we propose here 
to use {\it the same} basic phenomenological 
CD model of Miller et al.~(2004), simply allowing 
for the {\tt diskbb} normalization to assume 
negative as well as positive values.


For this paper, we choose to illustrate this issue 
with a fit to Holmberg II X-1, 
leaving a more extensive analysis to further work.
This is one of the sources in the 
Stobbart et al. (2006)'s sample for which a significant 
spectral slope change was reported ($>4$\,$\sigma$ improvement 
with respect to a single power-law fit); it is also 
one of the sources that appear to require an optically-thick, 
low-temperature corona 
when fitted with comptonization models (Goad at al.~2006). 
The {\it XMM-Newton} dataset studied by Goad at al.~(2006) and 
Stobbart et al. (2006) was from 2004 April 15. 
Instead, the data we discuss here were obtained from 
an earlier 9.8-ks {\it XMM-Newton} observation 
taken on 2002 April 16, during the historically 
highest state for this source (Dewangan et al.~2004).
To increase the signal-to-noise ratio, we coadded 
the EPIC pn and MOS data, using the code of Page, Davis 
\& Salvi (2003).  We fitted the source spectrum 
with a {\tt wabs} $\times$ {\tt tbvarabs} $\times$ 
({\tt diskbb} $+$ {\tt po}) model, to account for a 
(fixed) line-of-sight Galactic column as well as intrinsic, 
metal-poor absorption. We obtain a statistically good fit 
($\chi^2_\nu = 222.1/214 = 1.04$), 
as expected, with the following parameter values 
(Figure 2 and Table 1): intrinsic column density 
$N_{\rm H} = (1.6\pm0.1) \times 10^{21}$ cm$^{-2}$ (at a metal 
abundance $\approx 0.6 Z_{\odot}$), 
power-law slope $\Gamma = 2.38\pm0.06$, 
inner-disc temperature $kT_{\rm in} = 0.18 \pm 0.01$ keV, 
{\tt diskbb} normalization $K = 245^{+270}_{-145}$. Taken 
at face value, this suggests a mass 
$M (\cos \theta)^{1/2} = 630^{+280}_{-230} M_{\odot}$.

Hovever, we also obtain a good fit
($\chi^2_\nu = 222.3/214 = 1.04$), statistically 
indistinguishable from the other, with the following
parameter values (Figure 3 and Table 1):
intrinsic column density 
$N_{\rm H} = (1.9\pm0.1) \times 10^{21}$ cm$^{-2}$ (at a metal 
abundance $\approx 0.5 Z_{\odot}$), 
power-law slope $\Gamma = 2.67\pm0.09$, 
inner-disc temperature $kT_{\rm in} = 0.52 \pm 0.06$ keV, 
{\tt diskbb} normalization $K = -1.32^{+0.72}_{-1.27}$. Taken 
at face value, this suggests a mass 
$M (\cos \theta)^{1/2} = 45^{+19}_{-15}\, i M_{\odot}$, 
making it a robust iMBH (imaginary-mass BH) candidate.

Our preliminary investigation of other ULXs shows that 
a similar degeneracy is in fact common to many sources: 
in summary, a soft excess arises every time the power-law slope 
is constrained to fit the $2$--$5$ keV range, while 
a soft deficit is found when the slope fits 
the $5$--$10$ keV range. Either the excess or the deficit 
can then be accounted for, equally well, by adding 
or subtracting a {\tt diskbb} 
component. This degeneracy is a direct consequence 
of a change in slope at $\sim 5$\,keV in many bright ULXs, 
with the spectrum becoming steeper at higher energies 
(Stobbart et al.~2006). 
If all we are looking for, in this 
phenomenological model, is a simple, robust 
``common currency'' modelling for the comparison of different sources, 
there is no need to prefer the positive rather than 
the negative variety. In fact, if anything, 
the soft-deficit scenario is more consistent 
with the interpretation of bright ULXs 
as analogous to BHCs in the steep-power-law state, with 
additional absorption from heavily ionized metals 
around 1 keV. This would reconcile the power-law 
slope in ULXs with the typical indices seen 
in that spectral state for Galactic BHCs (Section 2.3).

Less phenomenological, more complex physical models 
would then show that the soft deficit is obviously not due 
to a negative {\tt diskbb} spectrum, but for example 
to smeared absorption lines. On the other hand, even 
the soft excess may not be true {\tt diskbb} emission, 
but instead caused by smeared emission lines and reflection.
In practice, the situation is likely to be even more 
complicated, with a basic underlying power-law spectrum 
modified by a mixture of emission, absorption 
and reflection.
Our definition of a soft excess or soft deficit 
is likely to have little absolute physical meaning, 
depending strongly on the fitting range and 
the detector sensitivity, in addition 
to real physical quantities.

\begin{figure}
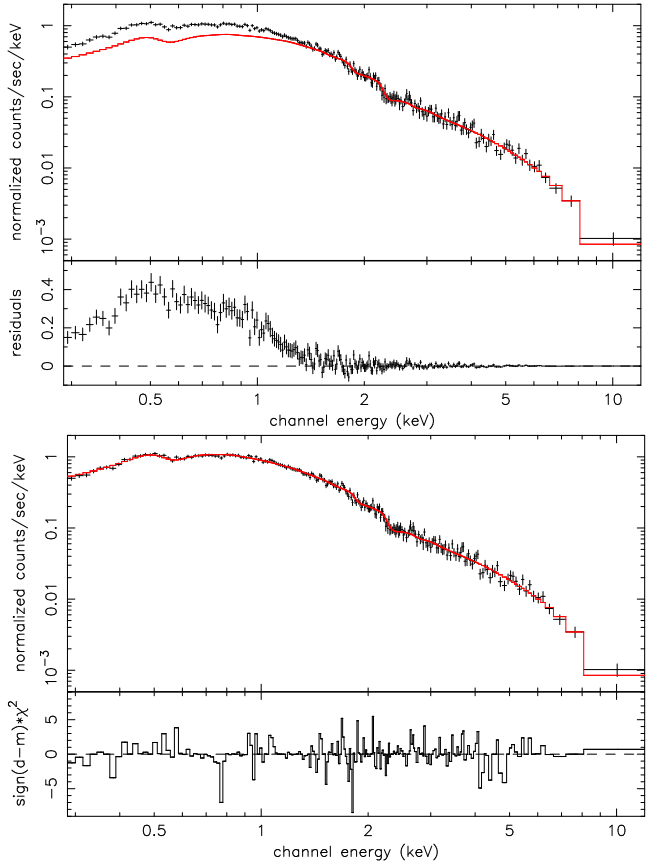

\epsfig{figure=ho2_softexcess.ps, width=5.7cm, angle=270}\\
\epsfig{figure=ho2_pos.ps, width=5.7cm, angle=270}
\caption{{\it XMM-Newton}/EPIC spectrum (coadded pn and MOS) 
and fit residuals for the Holmberg II X-1 ULX, 
observed on 2002 April 16 in 
its highest state (Dewangan et al.~2004).  We have fitted 
the spectrum with an absorbed power-law plus a {\it positive} 
disc-blackbody component (see Table 1 for the best-fit parameters). 
In the top panel, we have removed the disc-blackbody component 
to illustrate the amount of ``soft-excess'' above 
the power-law flux. Taken at face value,  
the {\tt diskbb} temperature and normalization 
provide robust evidence of an intermediate-mass BH (IMBH).}
\end{figure}

\begin{figure}
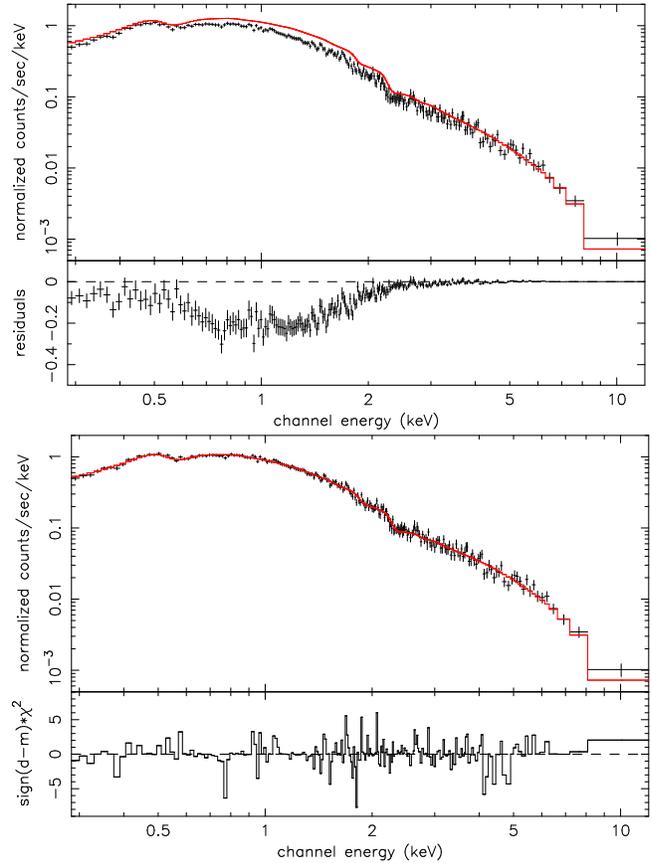

\epsfig{figure=ho2_softdeficit.ps, width=5.7cm, angle=270}\\
\epsfig{figure=ho2_neg.ps, width=5.7cm, angle=270}
\caption{{\it XMM-Newton}/EPIC spectrum (coadded pn and MOS) 
and fit residuals for the Holmberg II X-1 ULX, as in Figure 1. 
Again, we have fitted the spectrum with an absorbed power-law 
plus a disc-blackbody component 
(see Table 1 for the best-fit parameters).
However, this time we have allowed the {\tt diskbb} normalization
to be {\it negative}, obtaining an equally-good fit, 
but with a power-law slope more consistent with 
the values found in bright Galactic BHCs.
In the top panel, we have removed the disc-blackbody component 
to illustrate the amount of ``soft-deficit'' below 
the power-law flux. Taken at face value,  
the {\tt diskbb} temperature and normalization 
provide robust evidence of an imaginary-mass BH (iMBH).}
\end{figure}

\section{Blurred, ionized absorption and emission}

\subsection{Description of our model}

Complex physical models based on blurred emission and absorption 
lines and reflection   
were developed, amongst others, by Ross, Fabian \& Ballantyne
(2002), Gierlinsky \& Done (2004),  Crummy et al. (2006), and
Chevallier et al. (2006), initially to explain soft excess and
reflection bumps in AGN. Here we develop Chevallier et al.~(2006)'s 
model, based on the presence of highly ionized plasma in
the line-of-sight of the primary X-ray source, and apply it 
to the observed spectra of bright ULXs.

In our modelling, we assume a primary source of radiation
characterized by a power-law spectrum extending from $10$ 
to $10^{5}$ eV; we also assume that this primary emission is
produced close to the BH, and we do not speculate at this stage 
what physical mechanism is responsible for it. 
Any medium surrounding this 
primary source will be radiatively heated and photoionized. If
the ionized medium is sufficiently far from the ionizing source,
we can treat it in a 1-D plane-parallel geometry, as a slab of gas
illuminated from one side by a radiation field concentrated in a
very small, pencil-like, shape centered on the normal direction.
The resultant spectrum, reprocessed by such a medium, would be a
combination of ``reflection'' from the illuminated side (not a real
reflection, as it includes atomic and compton reprocessing),
``outward emission'' (coming from the non-illuminated side
of the medium), and a transmitted fraction of the incident 
ionizing continuum.
The relative contribution of each component 
to the total, observed spectrum depends on 
parameters such as the size, density and geometry of the ionized 
medium. Those parameters are directly related to the covering factor: 
if this factor is very close to unity, only the outward 
emission plus the transmitted (partly absorbed) 
component will be observed; if it is small, then the primary source 
spectrum can be  observed together with the reflection 
and outward emission (see fig. 1 in Chevallier et al. 2006). 

The models were computed using 
the photoionization code {\small TITAN} 
(Dumont, Abrassart \& Collin 2000; 
Dumont et al. 2002; Collin, Dumont \& Godet 2004), which is
well suited for the study of both optically thick (Thomson
optical depth up to several tens) and thin ionized media, 
such as warm absorbers (Gon\c{c}alves et al. 2006). 
Its advantage over other photoionization codes, such as 
{\small CLOUDY} (Ferland et al.~1998) or
{\small XSTAR} (Kallman \& Bautista~2001), is that it treats 
the transfer of both the lines and the continuum using 
the Accelerated Lambda Iteration ({\small ALI}) method (see 
Dumont et al. 2003 for a description of the {\small ALI} method 
in the modelling of the X-ray spectra of AGN and X-ray binaries); 
in addition, it can work under constant total
(gas plus radiation) pressure, thus offering a more adequate
treatment of the highly ionized gas in the vicinity of a
strong X-ray source (Gon\c{c}alves et al. 2006). 
{\small TITAN} includes all relevant physical processes 
from each level (e.g., photoionization, radiative and 
dielectronic recombination, ionization by high energy photons, 
fluorescence and Auger processes,  collisional ionization, 
radiative and collisional excitation/de-excitation, etc.) 
and all induced processes. It solves the ionization 
equilibrium of all the ion species of each 
element\footnote{Our atomic data include $\sim 10^{3}$ 
lines from ions and atoms of H, He,  C, N, O,  Ne,  Mg, Si, 
S, and Fe.}, the thermal equilibrium, the statistical 
equilibrium of all the levels of each ion, and the transfer 
of the lines and of the continuum. 
It gives as output the ionization, density and temperature 
structures, as well as the reflected and outward spectra.   
The energy balance is ensured locally with a precision of 
0.01\%, and globally with a precision of 1\%; as a consequence, 
the total reflected and outward components,  integrated over 
all solid angles, are constrained to be equal (within 1\%) to 
the primary ionizing spectrum, and the total output flux 
(leaving the gas slab from both sides) is the same as the 
injected flux.

All models were computed using the cosmic abundances of 
Allen (1973) and assuming the gas to be in total
pressure equilibrium. They were then convolved 
with a relativistic wind (Chevallier et al. 2006) 
with a dispersion velocity $v = 0.1c$, which has the effect 
of blurring all the emission and absorption features. 
We have built separate grids of photoionization models 
for the absorption, emission, and reflection components. 
The models in our grids are described by: the ionization parameter 
$\xi = L_{\rm i}/n_{\rm H}R^{2}$ (where $L_{\rm i}$ 
is the luminosity in the 0.01--100\,keV range, $n_{\rm H}$ 
is the hydrogen number density at the illuminated side of the 
medium, and $R$ is the distance from the ionized plasma to the
illuminating source); the column density $N^i_{\rm H}$ 
of the ionized gas; and the spectral energy
distribution of the incident X-ray flux, in our case 
a power-law spectrum parametrized by its photon 
index $\Gamma$. The ionized gas density $n_{\rm H}$ 
was set to $10^{11}$ cm$^{-3}$; however, the output spectrum 
depends only very weakly on this parameter, 
which simply rescales the spatial size of the ionized 
gas region, for given values of $\xi$ and $N^i_{\rm H}$.
Our grids cover the following parameter space:
$1000 \le \xi \le 4000$, $10^{22} \le N^i_{\rm H} \le 10^{23}$ 
and $2.4 \le \Gamma \le 3.3$.
Finally, the model grids were converted into  
additive table models ({\tt atable}) in {\small XSPEC}, allowing us 
to fit real data. Additional neutral absorption, 
from cold gas further away in the local ULX environment 
and along the line-of-sight in our Galaxy, 
was also added within {\small XSPEC}.
In practice, we found that the reflection 
component provides only a negligible improvement  
to our ULX fits, in agreement with a covering factor 
close to unity; therefore, only the absorption and 
emission components were taken into 
account, thus reducing the number of free parameters in 
our models.

\begin{figure}
\epsfig{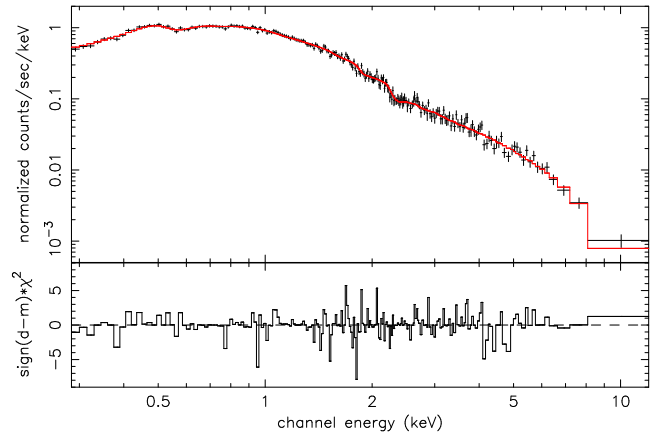}
\caption{{\it XMM-Newton}/EPIC spectrum (coadded pn and MOS) 
and fit residuals for the Holmberg II X-1 ULX, as in Figures 1, 2. 
Here, we have fitted the spectrum with a physical model 
based on an injection power-law spectrum, 
slightly modified by smeared emission and absorption 
lines from a slab of highly-ionized gas in front of 
the X-ray source. See Section 3 for details of the model, and Table 1 
for the best-fit parameters. In this scenario, the disc 
is not directly visible.}
\end{figure}

\subsection{Fitting to real data}

Once again, we fitted the {\it XMM-Newton}/EPIC spectrum of 
Holmberg II X-1 (Section 2.4), this time using 
our relativistically smeared ionized plasma models (labelled as 
{\tt Tabs} and {\tt Temi} in Table 1 and 2).
The best fit gives slightly better results  
($\chi^2_\nu = 213.5/213 = 1.00$) than the phenomenological  
{\tt diskbb} fits (Figure 4 and Table 1).
The most significant best-fit parameters are: 
an injection power-law spectrum with $\Gamma = 2.66\pm0.05$;
a neutral absorber column density 
$N_{\rm H} = (1.9 \pm 0.1) \times 10^{21}$ cm$^{-2}$ 
(in addition to the Galactic line-of-sight absorption), 
with metal abundance $Z \approx 0.5 Z_{\odot}$;
a column density $N^i_{\rm H} = (3.2 \pm 0.4) \times 10^{22}$ 
cm$^{-2}$ for the ionized plasma (ionization parameter 
$\xi = 2740\pm750$). As an aside, we also tried the HD model 
but it does not provide an acceptable fit to this source.

We then repeated the same exercise with the {\it XMM-Newton}/EPIC
spectrum of another bright ULX displaying a curvature at 
higher energies ($>3$\,$\sigma$ improvement by using a broken 
power-law fit with respect to a single power-law): NGC~4559 X-1. 
This source is often cited as an intermediate-mass BH candidate 
partly because of its cool soft-excess component 
(Cropper et al.~2004, where the source is identified 
as ``X-7'', using the old {\it ROSAT} identification).
In agreement with Cropper et al.~(2004), we find 
that interpreting that component as cool disc-blackbody 
emission (using a positive {\tt diskbb} or any of the various  
comptonization models in {\small XSPEC}) leads to estimated
masses $\sim 10^3 M_{\odot}$; the power-law index 
is $\Gamma \approx 2.2$ (Figure 5, top panel, and Table 2). 
However, as for Holmberg II X-1, 
a negative {\tt diskbb} component with a steeper power-law 
index ($\Gamma \approx 2.7$) provides an equally good (in fact, 
better) fit; taken at face value, the disc-blackbody 
normalization corresponds to a mass $\approx 100 i M_{\odot}$ 
(Figure 5, middle panel, and Table 2). 
For this source, the HD model also provides an acceptable fit 
to the data (Table 2); however, 
it is not as good as the alternative models.
Finally, we obtained a good fit using our 
ionized plasma model\footnote{In this case, the parameter 
$\xi$ pegged at its lowest value of $1000$; a better fit 
would have been obtained if our grids had extended 
to slightly lower values, which will be done in forthcoming 
works.}, with $\Gamma \approx 2.7$  
(Figure 5, bottom panel, and Table~2).

It is not surprising that our relativistically smeared, 
photoionized plasma models provide good fits: after all, 
they provide broad, smooth emission and absorption 
bumps from the reprocessing of an injected power-law spectrum. 
We already know that the spectra of bright ULXs 
can be fitted by a steeper power-law plus a broad absorption 
feature, or by a flatter power-law plus a broad emission feature, 
or by a combination of both. Here, we argue  
that such absorption and emission bumps are easily, 
self-consistently produced by an X-ray irradiated plasma 
(i.e., not by optically-thick accretion disc emission), 
under a plausible range of physical parameters, with the only 
more stringent condition that the lines be blurred 
by modest relativistic motions. 
The amount of blurring in our models (corresponding to a 
velocity dispersion $v/c= 0.1)$ has been 
assumed based on the observations, so that we get  
smooth continuum and bumps, with no narrow features 
(see Chevallier et al. 2006). 
Such high velocities could be due, for example, to fast outflows, 
or Keplerian motion very close to the BH.

\begin{table*}
\begin{tabular}{lcccc}
\hline
Selected parameters 
	& \multicolumn{4}{c}{Model: 
	{\tt wabs} $\times$ {\tt tbvarabs} $\times$} \\
\hline
 & ({\tt diskbb} $+$ {\tt po})$_{+}$ & 
   ({\tt diskbb} $+$ {\tt po})$_{-}$ & 
   {\tt bmc} & ({\tt Tabs} $+$ {\tt Tem})\\
\hline
$N_{\rm H}$ ($\times 10^{21}$ CGS) & $1.6^{+0.1}_{-0.1}$  & 
	$1.9^{+0.1}_{-0.1}$  & 
	$2.1^{+0.5}_{-0.4}$ & $1.9^{+0.1}_{-0.1}$ \\[3pt]
$Z/Z_{\odot}$    & $0.6^{+0.1}_{-0.2}$ & $0.5^{+0.2}_{-0.1}$ 
	 & $1.0^{+0.3}_{-0.3}$  & $0.5^{+0.3}_{-0.2}$ \\[3pt]
$\Gamma$    & $2.38^{+0.06}_{-0.06}$ & $2.67^{+0.09}_{-0.09}$ 
	 & $2.41^{+0.04}_{-0.04}$  & $2.66^{+0.05}_{-0.05}$ \\[3pt]
$kT_{\rm in}$ (keV)    & $0.18^{+0.01}_{-0.01}$  &
	$0.52^{+0.06}_{-0.06}$ 
	& $-$  & $-$ \\[3pt]
$K_{\rm dbb}$    & $245^{+270}_{-145}$ & $-1.32^{+0.72}_{-1.27}$   
	 & $-$  & $-$ \\[3pt]
$kT_{0}$ (keV)   & $-$ & $-$  & $0.13^{+0.01}_{-0.01}$  & $-$ \\[3pt]
$N^i_{\rm H}$ ($\times 10^{22}$ CGS)    & $-$ & $-$ & $-$  & 
	$3.2^{+0.4}_{-0.4}$ \\[3pt]
$\xi$    & $-$ & $-$ & $-$   & $2740^{+750}_{-750}$ \\
\hline
$\chi^2_{\nu}$    & $222.1/214$ & $222.3/214$   & 
	$231.2/214$  & $213.5/213$ \\
		    & $(1.04)$ & $(1.04)$  & $(1.08)$ & $(1.00)$ \\
\hline
$f^{\rm tot}_{0.3-10}$ ($\times 10^{-12}$ CGS)    & $12.4$ & $13.4$  &  
	$10.6$ & $13.2$  \\[3pt]
$f^{\rm po}_{0.3-10}$ ($\times 10^{-12}$ CGS)    & $9.6$ & $15.3$  &  $-$
	& $-$  \\[3pt]
$|f^{\rm abs}_{0.3-10}|$ ($\times 10^{-12}$ CGS)    & $-$ & $1.9$ &  $-$
	& $-$ \\[3pt]
$f^{\rm em}_{0.3-10}$ ($\times 10^{-12}$ CGS)    & $2.8$  & $-$  & $-$ 
	 & $-$ \\[3pt]
$L_{0.3-10}$ ($\times 10^{40}$ CGS)   & $1.8$ & $2.0$   &  $1.6$
	& $2.0$ \\[3pt]
\hline
$(M_{\rm BH}/M_{\odot}) \times(\cos \theta)^{1/2}$  
	& $630^{+280}_{-230}$  & $45^{+10}_{-15} i$ 
	 & $\approx 10^3$ & $-$ \\
\end{tabular}
\caption{Main best-fit parameters for the {\it XMM-Newton}/EPIC 
spectrum of Holmberg~II X-1, using a number of different 
phenomenological and physical models: phenomenological 
disc-blackbody models in emission and absorption 
(IMBH and iMBH, respectively), a comptonization model, 
and our ionized 
outflow model; the hot-disk model is not give here, as it does 
not provide an acceptable fit to this source. For {\tt diskbb} models, 
the characteristic mass 
$(M_{\rm BH}/M_{\odot}) \times(\cos \theta)^{1/2}$ 
is defined as 
$(K_{\rm dbb})^{1/2} \times (d/10{\rm pc}) \times [1/(6\times 1.5)]$.
We then list the total unabsorbed flux ($f^{\rm tot}_{0.3-10}$) 
and isotropic luminosity ($L_{0.3-10}$) in the $0.3$--$10$ keV band; 
for the {\tt diskbb} models, we also give the unabsorbed flux 
that appears added ($f^{\rm em}_{0.3-10}$) or subtracted 
($f^{\rm abs}_{0.3-10}$) to the power-law in the form 
of soft excess or soft deficit.
}\vspace{-2mm}
\end{table*}
\begin{table*}
\begin{tabular}{lccccc}
\hline
Selected parameters 
	& \multicolumn{5}{c}{Model: 
	{\tt wabs} $\times$ {\tt tbvarabs} $\times$} \\
\hline
 & ({\tt diskbb} $+$ {\tt po})$_{+}$ & 
   ({\tt diskbb} $+$ {\tt po})$_{-}$ & 
   ({\tt bb} $+$ {\tt diskbb}) & 
   {\tt bmc} & ({\tt Tabs} $+$ {\tt Tem})\\
\hline
$N_{\rm H}$ ($\times 10^{21}$ CGS) & $2.5^{+0.3}_{-0.3}$  & 
	$2.6^{+0.3}_{-0.3}$ & $0.9^{+0.1}_{-0.1}$ & 
	$2.1^{+0.5}_{-0.4}$ & $2.6^{+0.5}_{-0.4}$ \\[3pt]
$Z/Z_{\odot}$    & $0.5^{+0.2}_{-0.2}$ & $0.3^{+0.1}_{-0.1}$ & $<0.1$
	 & $0.5^{+0.3}_{-0.2}$  & $0.5^{+0.3}_{-0.2}$ \\[3pt]
$\Gamma$    & $2.24^{+0.05}_{-0.05}$ & $2.66^{+0.05}_{-0.05}$ & $-$
	 & $2.25^{+0.04}_{-0.04}$  & $2.73^{+0.10}_{-0.05}$ \\[3pt]
$kT_{\rm in}$ (keV)    & $0.14^{+0.01}_{-0.01}$  &
$0.42^{+0.03}_{-0.03}$ 
	& $1.35^{+0.06}_{-0.05}$ & $-$  & $-$ \\[3pt]
$K_{\rm dbb}$    & $160^{+340}_{-95}$ & $-0.80^{+0.35}_{-0.47}$  & 
	$(9.1^{+1.5}_{-1.7})\times 10^{-3}$ & $-$  & $-$ \\[3pt]
$kT_{0}$ (keV)   & $-$ & $-$  & $-$ & $0.11^{+0.01}_{-0.01}$  & $-$ \\[3pt]
$kT_{\rm bb}$ (keV)   & $-$ & $-$  & $0.18^{+0.01}_{-0.01}$ & $-$  & $-$ \\[3pt]
$K_{\rm bb}$   & $-$ & $-$  & $(4.4^{+0.01}_{-0.01})\times 10^{-6}$ 
	& $-$  & $-$ \\[3pt]
$N^i_{\rm H}$ ($\times 10^{22}$ CGS)    & $-$ & $-$ & $-$ & $-$ & 
	$3.7^{+0.2}_{-0.2}$ \\[3pt]
$\xi$    & $-$ & $-$ & $-$  & $-$ & $[1000]$ \\
\hline
$\chi^2_{\nu}$    & $216.7/208$ & $199.8/208$  & $229.6/208$ & 
	$215.2/208$  & $205.5/208$ \\
		    & $(1.04)$ & $(0.96)$  & $(1.10)$ & $(1.03)$ & $(0.99)$ \\
\hline
$f^{\rm tot}_{0.3-10}$ ($\times 10^{-12}$ CGS)    & $1.80$ & $1.76$  & $0.95$ &  
	$1.53$ & $2.01$  \\[3pt]
$f^{\rm po}_{0.3-10}$ ($\times 10^{-12}$ CGS)    & $1.19$ & $2.20$  & $-$ &  $-$
	& $-$  \\[3pt]
$|f^{\rm abs}_{0.3-10}|$ ($\times 10^{-12}$ CGS)    & $-$ & $0.44$ & $-$ &  $-$
	& $-$ \\[3pt]
$f^{\rm em}_{0.3-10}$ ($\times 10^{-12}$ CGS)    & $0.61$  & $-$  & $-$  & $-$
	 & $-$ \\[3pt]
$f^{\rm hd}_{0.3-10}$ ($\times 10^{-12}$ CGS)    & $-$  & $-$  & $0.62$ & $-$
	 & $-$ \\[3pt]
$f^{\rm bb}_{0.3-10}$ ($\times 10^{-12}$ CGS)    & $-$  & $-$  & $0.33$  & $-$
	 & $-$ \\[3pt]
$L_{0.3-10}$ ($\times 10^{40}$ CGS)   & $2.0$ & $2.0$ & $1.1$  &  $1.7$
	& $2.2$ \\[3pt]
\hline
$(M_{\rm BH}/M_{\odot}) \times(\cos \theta)^{1/2}$  
	& $1400^{+1600}_{-500}$  & $100^{+25}_{-25} i$  & $11^{+1}_{-1}$
	 & $\approx 10^3$ & $-$ \\
\end{tabular}
\caption{As in Table 1, for the {\it XMM-Newton}/EPIC 
spectrum of NGC\,4559 X-1. 
}
\end{table*}

\begin{figure}
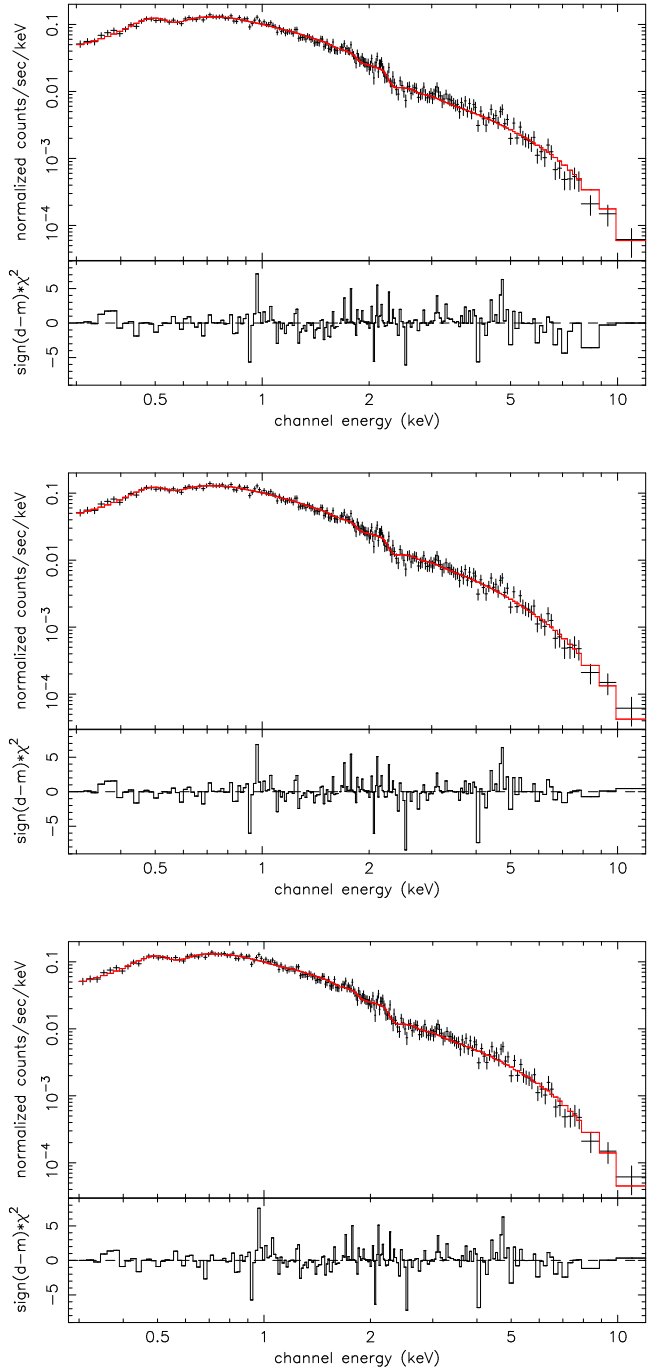

\epsfig{figure=ngc4559_diskbbpo_pos.ps, width=5.7cm, angle=270}\vspace{5mm}
\epsfig{figure=ngc4559_diskbbpo_neg.ps, width=5.7cm, angle=270}\vspace{5mm}
\epsfig{figure=ngc4559_absemi.ps, width=5.7cm, angle=270}\vspace{3mm}
\caption{Three statistically-good fits 
to the {\it XMM-Newton}/EPIC spectrum 
of NGC\,4559 X-1 (Cropper et al.~2004), 
with three different models. Top panel:
the spectrum is modelled with an underlying flatter 
power-law ($\Gamma \approx 2.2$) plus a soft excess, 
approximated by a (positive) {\tt diskbb} component at $0.14$ keV.
Middle panel: the spectrum is modelled with a steeper 
($\Gamma \approx 2.7$) power-law with a broad 
absorption feature approximated by a (negative) 
{\tt diskbb} component at $0.42$ keV. We argue that 
neither the positive nor the negative {\tt diskbb} 
component has any physical meaning or relation with 
the accretion disc; they are simply convenient, 
versatile components to model broad bumps. Bottom panel: 
the same spectrum, modelled with an underlying power-law 
($\Gamma \approx 2.7$) modified self-consistently 
by smeared emission and  
absorption lines caused by a layer of highly 
ionized gas. See Table 2 for the best-fit parameters of all 
three models.}
\end{figure}

\section{Summary and Discussion}

\subsection{The weakness of phenomenological models}

Phenomenological models can successfully reproduce 
the observed X-ray spectral features of ULXs, 
and may be useful for the classification and comparison 
of different classes of sources, but are not 
necessarily more constrained, robust or reliable than complex 
physical models. In fact, they often introduce a dangerous 
bias, when we try to attribute them unwarranted 
physical meaning. The disc-blackbody component often used 
to model deviations from a pure power-law spectrum 
in ULXs and AGN is a classical example. We argued that 
the so-called soft excess can just as easily be described 
as a soft deficit, depending on the energy range to which 
we choose to fit the power-law continuum. A small change 
of the fitted power-law slope can turn an apparent absorption 
feature into an apparent emission feature. Both the excess 
and the deficit are well modelled with a disc-blackbody component  
in emission or absorption, respectively. 
This is simply because a disc-blackbody component is 
a versatile tool to model a broad, smooth bump or trough 
(especially if we can adjust both the temperature and 
the absorbing column density), regardless of its original 
meaning of an accretion disc spectrum.
Similarly, it could be tempting to classify 
ULXs as stellar-mass objects, at least for those sources 
(see Stobbart et al.~2006) where a hot-disc model reproduces 
the observed curvature around $5$ keV better than a simple 
power-law. Again, we argued that this is misleading:  
the curvature is likely to be unrelated 
to a true disc spectrum, and may instead be the signature 
of increasing, smeared absorption at energies below $5$ keV. 

A more physical model to bright ULX spectra 
is likely to require an injected power-law 
spectrum modified by the presence of an 
ionized medium in our line-of-sight, resulting in 
a combination of blurred, 
blended emission and absorption lines, and possibly 
also reflection bumps. Such kind of models would 
give a more natural explanation for the reason  
why the X-ray spectra of many bright ULXs and 
soft-excess AGN look somewhat similar.
We have shown one recent implementation of such complex models, 
which we have developed thanks to the photoionization 
code {\small TITAN}, imported into {\small {XSPEC}},  
and applied to two bright ULXs as an illustrative example 
for this paper. We are aware of the larger number 
of free parameters included in this kind of models, 
but we argue that they are less 
misleading than {\tt diskbb} models. Our modelling shows 
that it is possible to produce broad, smooth emission 
and absorption features when an injected power-law  
spectrum is seen through a highly-ionized plasma 
with midly relativistic motion. This may be an ad hoc 
condition at this stage, but it is probably less problematic 
than attributing those features to accretion disc 
emission. We do not speculate at this stage 
the geometry of the ionized plasma; 
we merely point out 
that such intervening medium would produce 
an effect consistent with what is observed. 
In principle, the relative contribution of 
the outward absorption and emission components to the  
observed spectrum could help us constrain the geometry 
of the ionized medium. However, this is not yet possible
with the available X-ray data, because the true slope 
of the injected power-law component is not known 
a priori and cannot be precisely determined over 
the small energy range of our detectors. 
Further work on a more extended sample of ULXs 
(Soria et al., in preparation) will begin to explore 
this issue, at least on a statistical level. 
Future observations with instruments such as 
the Hard X-ray Telescope on {\it Constellation-X}
will be needed to constrain the slope of the primary 
continuum over a larger energy range, and thus better 
determine the relative contribution of each components 
and the physical origin of the power-law itself.

The uncritical use of a disc-blackbody model, 
interpreted as robust evidence of cool disc emission, 
has led to claims of BH masses $\sim 10^3 M_{\odot}$, 
skewing both observational and theoretical studies 
of ULXs towards the IMBH scenario.
The ionized-plasma model does not 
provide in itself evidence in favour or against IMBHs (in fact, 
it can also be applied to IMBHs and to even bigger BHs, in AGN); 
however, it implies that the deviations from a power-law 
spectrum seen in bright ULXs are not related to disc emission, 
and therefore are not a measure of their BH masses. 
Without this piece of information, the remaining evidence 
in favour of IMBHs is much weakened 
for the majority of ULXs (except for the brightest source in M\,82). 
Time-variability studies may provide a more critical test   
(e.g., Markowitz et al.~2003; Markowitz \& Edelson 2004; 
Fiorito \& Titarchuk 2004;
Uttley \& McHardy~2005; Done \& Gierlinski 2005), at least for 
those few sources for which we have enough counts 
to investigate high-frequency quasi-periodic oscillations 
and breaks in the power density spectrum.
In particular, the temporal behaviour of Holmberg II X-1
was studied by Goad et al. (2006), who found substantial 
variability on time-scales of months to years, but very little 
variability on time-scales of less than a day. 
Based on a combination of energy spectrum and power spectral density 
considerations, they concluded that Holmberg II X-1 
was not likely to be in the disc-dominated
high/soft state, and may instead have been 
in the steep-power-law state (McClintock \& Remillard 2006);
this would be consistent with a BH mass $\la 100 M_{\odot}$. 
For NGC\,4559 X-1, a break in the power density spectrum 
was noted and discussed by Cropper et al.~(2004). 
However, they concluded that the time-variability 
data available are not yet sufficient to constrain the mass range 
significantly: BH masses $\sim 50 M_{\odot}$ 
or $\sim 1000 M_{\odot}$ could both be consistent 
with the observed break frequency, 
if one takes different assumptions on the interpretation 
of that break and its scaling with BH mass. 
For most other ULXs, the strongest constraint to their BH mass  
remains their X-ray luminosity in comparison with the Eddington 
limit. This argument suggests an upper limit 
$\sim 100$--$200 M_{\odot}$ if the emission is isotropic 
(Swartz et al.~2004; Gilfanov, Grimm \& Sunyaev 2004) 
and even less if beamed.
This may still be an order of magnitude higher 
than the mass of Galactic BHs, but may be accommodated 
with more ordinary star-formation processes.

\subsection{ULXs as a new spectral state?}

Complex models based on ionized absorption, emission and 
reflection assume that the underlying X-ray spectrum 
is a power-law, only slightly modified. 
This is also consistent with the detection 
of some bright ULXs with a pure power-law spectrum. 
How can the disc be not visible at all, not even 
as a small bump? We have argued that 
the brightness of those sources strongly disfavours 
models in which the disc is simply truncated far from the 
innermost stable orbit. A possible qualitative 
alternative is that the disc emission 
is completely comptonized in a non-thermal corona. 
This requires that most of the available 
gravitational power be efficiently released 
in the corona, or efficiently transferred 
from the disc to the corona---in this scenario, 
a magnetic disc/corona coupling could perhaps 
drain most of the energy from the disc 
at small radii (e.g., Kuncic \& Bicknell 2004).

A related problem is to determine the low-energy 
cut-off to the power-law which, we know for certain,  
does not extend to the optical band. In comptonization models, 
the low-energy cut-off identifies the characteristic temperature 
(or at least the lowest temperature) of the seed photons. 
In the ionized outflow model, the injected power-law 
component is assumed to extend without a break 
even below the {\it Chandra} or {\it XMM-Newton} 
energy band. This is also the case if we adopt 
a phenomenological disc-blackbody plus power-law model. 
Moreover, it appears to be the case for those bright ULXs 
that can be fitted with a simple power-law. 
Taken at face value, this requires the 
presence of seed photons at energies $\la 0.1$ keV: 
we speculate that UV photons from the outer disc and/or 
the donor star may also contribute as seed for the comptonization 
process. The fraction of such photons that may 
illuminate the comptonizing region and be upscattered   
depends on unknown parameters such as the radial and vertical 
size of the corona, the thickness and flaring 
angle of the disk, the spectral type and radius 
of the donor star, the binary separation.

A similar situation (i.e., dominant power-law component 
extending to energies $\la 0.3$ keV and comparatively 
small disc component) appears to occur in the steep-power-law 
state of Galactic BHCs (McClintock \& Remillard 2006), and 
is not well understood in that case, either. Also, in that 
state, the disc contribution is already small compared to 
the power-law. Goad et al. (2006) suggested that  
the temporal variability of Holmberg II X-1 is 
similar to that found in the Galactic BHC GRS~1915$+$105 in 
its steep-power-law state ($\chi$ class). Thus, we speculate that 
ULXs represent a further spectral state, contiguous to the 
steep-power-law state, in which 
the disc contribution is entirely negligible 
and, in addition, the dominant power-law component 
is modified by smeared emission and absorption 
from the surrounding, highly-ionized, possibly 
outflowing gas. Interestingly, one of the effects 
of the broad absorption features at $\sim 1$ keV 
is to make the continuum appear flatter 
than the injected power-law, 
over the $2$--$10$ keV range, as we noted 
when comparing positive and negative disc-blackbody 
models (see also fig.~9 in Chevallier et al.~2006).
This may be one reason why many bright ULXs in this class 
appear to have a flatter power-law slope, when fitted
with a CD model, than Galactic BHCs in the steep-power-law 
state (the latter presumably being less affected 
by highly-ionized, fast outflowing plasma). 

Such a spectral state could be shared by higher-mass 
accretors such as AGN. Narrow Line Seyfert 1s, in particular, 
display a soft X-ray excess and characteristic variability which 
could be associated with a steep-power-law state. 
It has been shown (Chevallier et al.~2006) that the soft 
excess in AGN could be fitted with the same relativistically 
smeared ionized plasma model applied here to ULXs. Thus, our approach 
offers a possible common explanation to the properties 
of ULXs, soft-excess AGN and Galactic BHs; 
it suggests that the main spectral features in this 
bright state depend on the physical parameters of 
the outflowing plasma, not on the mass of the accretor. 
A more detailed discussion of this issue is beyond the scope 
of this work; we shall address it in a forthcoming 
paper.



\section*{Acknowledgements}
We thank Zdenka Kuncic for discussions on the disk/corona 
coupling and Mat Page for pointing out the usefulness 
of combining EPIC spectra, and for letting us use his code.
We also thank the anonymous referee for insightful comments 
and comparisons with previous work.
ACG acknowledges support from the 
{\it Funda\c{c}\~ao para a Ci\^encia e a Tecnologia (FCT)}, 
Portugal, under grant BPD/11641/2002. RS acknowledges support 
from an OIF Marie Curie Fellowship, through University College London.

\end{document}